\documentclass[11pt]{amsart}

\usepackage{a4wide}
\usepackage{bbm}
\usepackage{amsxtra} 
\usepackage{t1enc} 
\usepackage{color}
\usepackage{epsf}
\usepackage{amssymb}
\usepackage{graphicx}
\usepackage{stmaryrd}

\DeclareMathAlphabet{\mathpzc}{OT1}{pzc}{m}{it} 

\newcommand{\N}{\mathbbm{N}}
\newcommand{\Z}{\mathbbm{Z}}
\newcommand{\Q}{\mathbbm{Q}}
\newcommand{\R}{\mathbbm{R}}
\newcommand{\C}{\mathbbm{C}}

\newtheorem{theorem}{Theorem}

\newtheorem{proposition}{Proposition}

\theoremstyle{definition}

\newtheorem{remark}{Remark}

\begin{document}

\title[Discrete tomography]{Discrete Tomography of Penrose Model Sets}

\author[M. Baake]{Michael Baake}
\author[C. Huck]{Christian Huck}
\address{\hspace*{-1em} Fakult\"{a}t f\"{u}r Mathematik,
  Universit\"{a}t Bielefeld, Postfach 10 01 31, 33501 Bielefeld, Germany}
\email{\{mbaake,huck\}@math.uni-bielefeld.de}
\urladdr{http://www.math.uni-bielefeld.de/baake/}

\begin{abstract} 
Various theoretical and algorithmic aspects of inverse problems in
discrete tomography of planar Penrose model sets are discussed. These are
motivated by the demand of materials science for the reconstruction of
quasicrystalline structures from a small number of images produced by quantitative high
resolution transmission electron microscopy.
\end{abstract}

\maketitle

\section{Introduction}
{\em Discrete tomography} is concerned with the 
inverse problem of retrieving information about some discrete object from 
(generally noisy) information about its incidences with certain query sets.
A typical example is the {\em reconstruction} of a finite planar point set
from its line sums in a small number of directions. More precisely,
for a {\em direction} $u\in\mathbb{S}^1$ (the unit circle), the ({\em discrete parallel}\/) {\em
  X-ray} $X_{u}F$ of a finite subset $F$ of
the Euclidean plane $\R^2$ in direction $u$ gives the
number of points of the set on each line in $\R^2$ parallel to
$u$, i.e., $X_{u}F$ is the function $X_{u}F: \mathcal{L}_{u}
\longrightarrow \mathbbm{N}_{0}:=\mathbbm{N} \cup\{0\}$, defined by
$$X_{u}F(\ell) := \operatorname{card}(F \cap \ell\,) =\sum_{x\in \ell}
\mathbbm{1}_{F}(x)\,,$$ where $\mathcal{L}_{u}$ is the set of lines in
direction $u$ in $\mathbbm{R}^{2}$, with obvious generalization to
higher dimensions. In the
classical setting, motivated by crystals, the positions to be
determined form a subset of a translate of the square lattice $\Z^2$ or, more
generally, of arbitrary lattices $L$ in $\R^d$, where $d\geq 2$. The
cases $d=2$ and $d=3$ are practically relevant. In fact, many of the problems in discrete tomography
have been studied on $\Z^2$, the `classical planar setting' of
discrete tomography; see~\cite{HK,Gr,GGP,GG}. 

In the longer run, by
also having other structures than perfect crystals in mind, one has to
take into account wider classes of sets, or at least
significant deviations from the lattice structure. As an intermediate
step between periodic and random (or amorphous) {\em Delone sets} (defined below), we consider systems of {\em aperiodic
  order}, more precisely, of so-called {\em model sets} (or {\em
  mathematical quasicrystals}), which are commonly accepted as a reasonable mathematical model for quasicrystalline structures in
nature~\cite{St}.

The main motivation for our interest in the discrete tomography of
model sets comes from the question how to reconstruct three-dimensional (quasi)crystals, or planar
layers of them, from their images under quantitative {\em high
  resolution transmission electron microscopy} (HRTEM) in a small
number of directions. In fact, in~\cite{ks,sk}, a technique called QUANTITEM
(quantitative analysis of the information coming from transmission
electron microscopy) is described, based
on HRTEM, which can effectively measure the number of atoms lying on
lines parallel to certain directions. In particular, with the growing importance of
surface science, there is also increasing interest in additional
methods for the reconstruction of planar structures which can
supplement STM approaches. At present, the measurement of the number
of atoms lying on a line can only be achieved for some crystals;
see~\cite{ks,sk}. However, it is reasonable to expect that future
developments in technology will improve this situation. It seems thus
timely to investigate the mathematical foundations now. 

Here, we restrict ourselves to an example, namely to the well-known class of planar model
sets $\varLambda_{\text{P}} \subset \R^2$ that are associated with the well-known
{\em Penrose tiling}, and present some results on the discrete tomography of these sets, with an
emphasis both on reconstruction and uniqueness problems. Note that 
proofs are omitted; details and extensions will appear in~\cite{H2,H3}. 

\section{Penrose Model Sets}\label{Penrose}

We always let $\zeta_{5}=e^{2\pi i/5}$, a primitive
 $5$th root of unity in $\C$. Then, $\Q(\zeta_{5})$ is the
corresponding 
cyclotomic field, an algebraic number field of degree $4$ over $\Q$, and $\Z[\zeta_{5}]$ is its subring of cyclotomic
integers. 

\begin{remark}
Let $\mathsf{C}_{5}$ denote the cyclic group of order $5$, i.e., $\mathsf{C}_{5}=\Z
/5\Z$. Moreover, $\mathsf{C}_{5}$ is understood to be supplied with the
discrete topology. It is well-known that every $z\in\Z[\zeta_{5}]$ can
uniquely be
written as 
$
z=\sum_{j=0}^{3}a_{j}(z)\zeta_{5}^{j}
$, 
 with $a_{j}(z)\in\Z$. Let
$\sigma_{2}$ be the automorphism of the Galois extension
$\Q(\zeta_{5})/ \mathbbm{Q}$ that is given by
$
\zeta_{5}\longmapsto \zeta_{5}^{2}
$.  
 Identifying $\R^2$ and $\C$ in the canonical way, $\sigma_{2}$ gives rise to a map 
$$
.\widetilde{\hphantom{a}}\,\,:\, \Z[\zeta_{5}] \longrightarrow
\R^{2}\times(\R^2\times \mathsf{C}_{5})\,,$$ defined by
$
z\longmapsto \Big(z,\big(\sigma_{2}(z),\sum_{j=0}^{3}(a_{j}(z)\;(\operatorname{mod}
    5))\big)\Big)$.
Via projection on the second factor, this induces a map
$
.^{\star}\,\,:\, \Z[\zeta_{5}] \longrightarrow
\R^2\times \mathsf{C}_{5}$, defined by
$
z\longmapsto \big(\sigma_{2}(z),\sum_{j=0}^{3}(a_{j}(z)\;(\operatorname{mod}
    5))\big)$. Then, $\Z[\zeta_{5}]\widetilde{\hphantom{a}}$ is a lattice in
$\R^{2}\times(\R^2\times \mathsf{C}_{5})$, i.e., a co-compact discrete
subgroup. In fact, $\Z[\zeta_{5}]\widetilde{\hphantom{a}}$ is the $\Z$-span of
the set $
\{1\widetilde{\hphantom{a}},(\zeta_{5})\widetilde{\hphantom{a}},(\zeta_{5}^2)\widetilde{\hphantom{a}},(\zeta_{5}^3)\widetilde{\hphantom{a}}\}$. Finally, note that $\Z[\zeta_{5}]^{\star}$ is dense in
$\R^2\times \mathsf{C}_{5}$; see~\cite{PABP}.
\end{remark}

It is well known by now that {\em model sets} arise from so-called {\em cut and
  project schemes}, compare~\cite{Moody,B}. In particular, the class of
{\em Penrose model sets} (PMS) arises from the following {\em cut and
  project scheme}; cf.~\cite{BM2}:

\begin{equation} \label{cppenrose}
\renewcommand{\arraystretch}{1.5}
\begin{array}{ccccc}
& \pi & & \pi_{\textnormal{\tiny int}}^{} & \vspace*{-1.8ex} \\
\!\!\!\!\!\!\!\!\!\!\!\!\!\!\!\R^{2} & \longleftarrow &
\;\;\;\;\;\R^{2}\times(\R^2\times \mathsf{C}_{5})  & \longrightarrow & \R^2\times \mathsf{C}_{5} \\
\!\!\cup \mbox{\tiny \, dense }&&\;\;\,\cup\mbox{\tiny\,
  lattice}&&\;\;\;\;\;\;\;\;\cup \mbox{\tiny \; dense}\\
 & \mbox{\tiny 1--1} & & \mbox{\tiny 1--1} & \vspace*{-3.4ex} \\
\!\!\!\!\!\!\!\!\!\!\!\!\!\!\! \Z[\zeta_{5}]&
\longleftrightarrow &
\underbrace{\Big \{\Big(z=\sigma_{1}(z),\underbrace{\big(\sigma_{2}(z),\sum_{j=0}^{3}(a_{j}(z)\;(\operatorname{mod}
    5))\big)}_{=z^{\star}}\Big)\,|\, z\in \Z[\zeta_{5}]\Big \} }_{=\Z[\zeta_{5}]\widetilde{\hphantom{a}}}& \longleftrightarrow &\,\,\,\,\Z[\zeta_{5}]^{\star} \\
\end{array}
\end{equation}

Given any subset $W\subset \R^2\times \mathsf{C}_{5}$ with
$\varnothing\, \neq\, W^{\circ}\subset W\subset \overline{W^{\circ}}$
and $\overline{W^{\circ}}$ compact, a so-called {\em window}, and any $t\in\R^2$, we obtain a
planar model set $\varLambda(t,W) := t+\varLambda(W)$ relative
to the above cut and project scheme~(\ref{cppenrose}) by setting $\varLambda(W):=\{z\in\Z[\zeta_{5}]\,|\,z^{\star}\in W\}$.

Let $P$ be the convex hull of the $5$th roots of unity, which is a regular pentagon centred at the origin. Set
$W^{(1)}:=P$, $W^{(2)}:=-P$, $W^{(3)}:=\tau P$ and
$W^{(4)}:=-\tau P$, with $\tau=(1+\sqrt{5})/2$ the golden ratio, and
$$
W_{P}:=\bigcup_{j=1}^{4} \left(W^{(j)}\times \{j\;(\operatorname{mod} 5)\}\right)\,\,\subset\,\, \R^2\times \mathsf{C}_{5}\,.
$$
Moreover, for $u\in \R^2$, set 
$W_{P}^{u}:=(u,0\;(\operatorname{mod}
5))+W_{P}$,
$
(W_{P}^{u})^{(j)}:=(u,0\;(\operatorname{mod}
5))+\big(W^{(j)}\times \{j\;(\operatorname{mod} 5)\}\big)
$
and
$\varLambda_{\text{P}}^{u}:= \varLambda(W_{P}^{u}) 
$. 
If $\varLambda_{\text{P}}^{u}$ is {\em generic}, i.e., if one has
$W_{P}^{u}\cap \Z[\zeta_{5}]^{\star}=\varnothing$, then all translates
 of
$\varLambda_{\text{P}}^{u}$, meaning the sets
$t+\varLambda_{\text{P}}^{u}$ with $t\in \R^2$, are called {\em
  Penrose model sets}. Note that this formulation avoids the usual
ambiguities from non-minimal embeddings into $5$-space.

\begin{figure}
\centerline{\epsfysize=0.50\textwidth\epsfbox{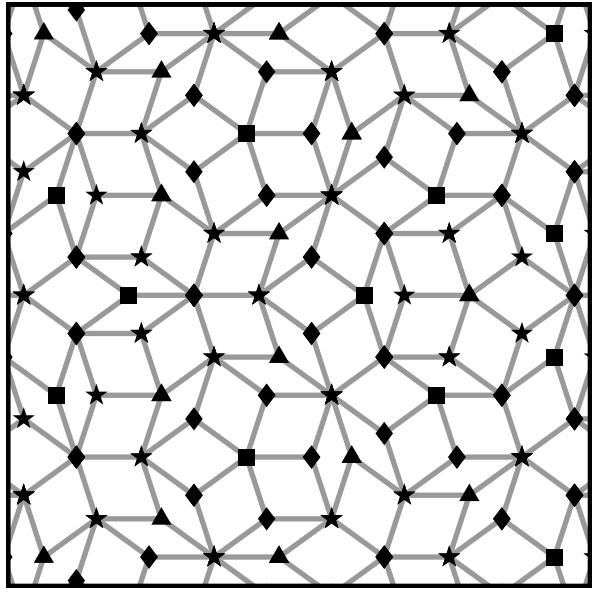}}
\caption{A patch of the fivefold symmetric Penrose tiling, with the four distinct classes of vertices indicated.}
\label{penfigure}
\end{figure}

\begin{remark}
$\varLambda_{\text{P}}^{0}$ is not generic, while
generic examples are obtained by shifting the window, i.e.,
$\varLambda_{\text{P}}^{u}$ is generic for almost all $u\in
\R^{2}$. Joining any two points with distance $1$ in a generic
$\varLambda_{\text{P}}^{u}$ by edges results in a
  {\em Penrose tiling}, which is a tiling with two types of
rhombi. See Figure~\ref{penfigure} for a generic example; different
generic choices of $u$ result in locally indistinguishable (LI)
Penrose tilings. Note that Penrose model sets
$\varLambda_{\text{P}}\subset\R^2$ are aperiodic, meaning that they
have no translational symmetries. Further, Penrose model sets are {\em
  Delone sets}, i.e.,
they are uniformly discrete and relatively
dense; cf.~\cite{Moody}.
\end{remark}

\section{Problems in Discrete Tomography of Penrose Model Sets}

Let $\varLambda_{\text{P}}$ be a PMS, $k\in\N$ and $R>0$. A finite subset $C$ of $\varLambda_{\text{P}}$ is called a {\em convex
set in} $\varLambda_{\text{P}}$ when its convex hull contains no new
points of $\varLambda_{\text{P}}$, i.e., when one has
$C=\operatorname{conv}(C)\cap\varLambda_{\text{P}}$. We
denote by $\mathcal{F}(\varLambda_{\text{P}})$, $\mathcal{F}_{\leq
  k}(\varLambda_{\text{P}})$, 
$\mathcal{D}_{<R}(\varLambda_{\text{P}})$ and
$\mathcal{C}(\varLambda_{\text{P}})$ the set of finite
subsets of $\varLambda_{\text{P}}$, the set of finite
subsets of $\varLambda_{\text{P}}$ having cardinality $\leq k$, the set of subsets of
$\varLambda_{\text{P}}$ with diameter less than $R$ and the set of convex
subsets of $\varLambda_{\text{P}}$, respectively.

\begin{remark}
The uniform
discreteness of Penrose model sets $\varLambda_{\text{P}}$ immediately implies the inclusion $\mathcal{D}_{<
  R}(\varLambda_{\text{P}})\subset \mathcal{F}(\varLambda_{\text{P}})$.
\end{remark}

Clearly, in order to obtain electron microscopy images of high
resolution, one should allow only directions 
which yield dense lines in Penrose model sets. These directions are clearly contained
in the set of all directions, called
$\varLambda_{\text{P}}${\em -directions}, which are parallel to a non-zero element
of the difference set of $\varLambda_{\text{P}}$, i.e.,  
$$\varLambda_{\text{P}}\!-\!\varLambda_{\text{P}}:=\{\lambda-\lambda'\,|\,\lambda,\lambda'\in\varLambda_{\text{P}}\}\,\,\subset \,\,
\Z[\zeta_{5}]\,.$$
 Calling a direction $u\in\mathbb{S}^1$ a $\Z[\zeta_{5}]${\em -direction} when it
is parallel to an element of $\Z[\zeta_{5}]\setminus\{0\}$, one has
the following result.
 
\begin{proposition}
If $\varLambda_{\textnormal{P}}$ is a \textnormal{PMS}, the  set of
$\varLambda_{\textnormal{P}}$-directions equals
the set of 
$\Z[\zeta_{5}]$-directions.
\end{proposition}

Let us indicate the main {\em algorithmic problems} in discrete tomography of Penrose model
sets. For a direction
$u\in\mathbb{S}^1$, we use 
$\mathcal{L}^{\Z[\zeta_{5}]}_{u}$ to denote the set of elements $\ell_{u}$ of
$\mathcal{L}_{u}$ that pass through a point of $\Z[\zeta_{5}]$. Let $u_1,\dots,u_m\in \mathbb{S}^1$ be $m\geq 2$ pairwise non-parallel
$\Z[\zeta_{5}]$-directions. The corresponding {\em consistency}, {\em reconstruction} and
{\em uniqueness problems} are defined as follows. 

\begin{quote}
{\sc Consistency}. \\
Given functions $p_{u_{i}}
  : \mathcal{L}_{u_{i}} \longrightarrow \mathbbm{N}_{0}$,
  $i\in\{1,\dots,m\}$, whose supports are finite and satisfy
  $\operatorname{supp}(p_{u_{i}})\subset
  \mathcal{L}^{\Z[\zeta_{5}]}_{u_{i}}$, decide whether there is a
finite set $F$ which is contained in a
 PMS and satisfies
$X_{u_{i}}F=p_{u_{i}}$, for all $i\in\{1,\dots,m\}$.  
\end{quote}

\begin{quote}
{\sc Reconstruction}. \\
Given functions $p_{u_{i}}
  : \mathcal{L}_{u_{i}} \longrightarrow \mathbbm{N}_{0}$,
  $i\in\{1,\dots,m\}$, whose supports are finite and satisfy
  $\operatorname{supp}(p_{u_{i}})\subset
  \mathcal{L}^{\Z[\zeta_{5}]}_{u_{i}}$; in the case that {\sc
    Consistency} is satisfied, construct 
a finite set $F$ which is contained in a
 PMS and satisfies $X_{u_{i}}F=p_{u_{i}}$, for all $i\in\{1,\dots,m\}$. 
\end{quote}

\begin{quote}
{\sc Uniqueness}. \\
Given a finite subset $F$ of a PMS, decide
whether there is a different finite set $F'$ that is also a subset of a
 PMS and satisfies
$X_{u_{i}}F=X_{u_{i}}F'$, for all $i\in\{1,\dots,m\}$.  
\end{quote}

In general, the above problem {\sc Reconstruction} can have many solutions
of rather different shape. Therefore, one is also interested in
{\em uniqueness results}, e.g., the
(unique) {\em determination} of the set 
$$
\bigcup_{\varLambda_{\text{P}}\mbox{\tiny \, PMS }}\mathcal{F}(\varLambda_{\text{P}})\,,
$$
or suitable subsets thereof by the $X$-rays in a small number of
$\Z[\zeta_{5}]$-directions. More precisely, we define the concept of
{\em determination} and the interactive concept of {\em successive
  determination} as follows. Let $\mathcal{E}$ be a collection of finite subsets of $\R^{2}$ and let $U\subset\mathbb{S}^1$ be a
finite set of directions. We say that $\mathcal{E}$ is {\em determined} by the $X$-rays in the directions of $U$ if, for all $F,F' \in \mathcal{E}$, one has
$$
(X_{u}F=X_{u}F',\;\,\,\forall u \in U) \;  \Longrightarrow\; F=F'\,.
$$
We say that $\mathcal{E}$ is {\em successively determined} by the
$X$-rays in the directions of $U=\{u_1,\ldots,u_{m}\}$, if,
for a given $F\in \mathcal{E}$, these can be chosen inductively (i.e., the choice of $u_{j}$ depends on all $X_{u_{k}}F$ with $k\in\{1,\ldots,j-1\}$) such that, for all $F' \in \mathcal{E}$, one has
$$
(X_{u}F'=X_{u}F,\;\,\,\forall u \in U)\;  \Longrightarrow\; F'=F\,.
$$
We say that $\mathcal{E}$ is {\em determined} (resp., {\em successively determined}\,) by $m$ $X$-rays if there is a set $U$ of $m$ pairwise non-parallel directions such that $\mathcal{E}$ is determined (resp., successively determined) by the $X$-rays in the directions of $U$. 
 
\section{Computational Complexity and Uniqueness Results}

Let us begin with a result on computational complexity, where we apply the real RAM-model of computation, see~\cite{PS}. Here, each of the standard elementary
operations on reals counts only with unit cost. This leads to the following tractability result.

\begin{theorem}\label{th2}
When restricted to two $\Z[\zeta_{5}]$-directions, the problems
{\sc Consistency}, {\sc Reconstruction} and {\sc Uniqueness} can be solved in polynomial time
in the real RAM-model.
\end{theorem}

\begin{remark}
It seems to be rather obvious from the results in~\cite{GGP} that one cannot expect a generalization of Theorem~\ref{th2} to the case of 
three or more $\Z[\zeta_{5}]$-directions. More precisely, we expect 
that, when restricted to three or more $\Z[\zeta_{5}]$-directions, the problems
{\sc Consistency}, {\sc Reconstruction} and {\sc Uniqueness} are $\mathbbm{NP}$-hard.
\end{remark}

Let us now present results dealing with the
(successive) determination of finite subsets of Penrose model
sets. Though we are not interested in
non-$\Z[\zeta_{5}]$-directions for practical reasons, we begin
with the following observation.

\begin{proposition}
If 
$\varLambda_{\textnormal{P}}$ is a \textnormal{PMS} and $u \in \mathbb{S}^{1}$ is a
non-$\Z[\zeta_{5}]$-direction, the class of finite subsets $\mathcal{F}(\varLambda_{\textnormal{P}})$ is determined by the single $X$-ray in direction $u$.
\end{proposition}

This last result immediately follows from the observation that, for all
PMS $\varLambda_{\text{P}}$, each line
in the plane in a non-$\Z[\zeta_{5}]$-direction passes through at most
one point of $\varLambda_{\text{P}}$, the latter being the reason for
the practical irrelevance of this result. On the other hand, the next result shows that any
fixed finite number of $X$-rays in $\Z[\zeta_{5}]$-directions does not
suffice to determine the whole class of finite subsets of a
fixed PMS $\varLambda_{\text{P}}$.

\begin{proposition}\label{source}
Let $\varLambda_{\textnormal{P}}$ be a \textnormal{PMS} and $U\subset
\mathbb{S}^{1}$ an arbitrary, but fixed finite set of pairwise
non-parallel $\Z[\zeta_{5}]$-directions. Then, the set $\mathcal{F}(\varLambda_{\textnormal{P}})$ is not determined by the $X$-rays in the directions of $\,U$.
\end{proposition}

In order to obtain results on uniqueness, one has to restrict the class
of finite sets under consideration. Within the class of finite subsets of a
fixed PMS $\varLambda_{\text{P}}$ with bounded
cardinality, there is the following result.

\begin{proposition}\label{m+1}
Let $\varLambda_{\textnormal{P}}$ be a \textnormal{PMS} and $k\in \mathbbm{N}$. Then, the
set 
$\mathcal{F}_{\leq k}(\varLambda_{\textnormal{P}})$ is determined by any set
of $k+1$ pairwise non-parallel $\Z[\zeta_{5}]$-directions, while any
set of 
$1+\lfloor\log_{2}k\rfloor$ pairwise non-parallel $X$-rays in
$\Z[\zeta_{5}]$-directions is insufficient for this purpose.
\end{proposition}

This last result is once again of limited relevance in practice, because
typical atomic structures to be determined comprise about $10^6$ to
$10^9$ atoms, and, in order not to damage or even destroy the examined
structures, one has to make sure that one uses at most $4$ or $5$ $X$-rays. 

\begin{proposition}\label{coromod}
Let $\varLambda_{\textnormal{P}}$ be a \textnormal{PMS} and $R>0$. Then, the set $\mathcal{D}_{<R}(\varLambda_{\textnormal{P}})$ is determined by two $X$-rays in $\Z[\zeta_{5}]$-directions.
\end{proposition}

Though the last result seems to be more satisfactory, it is probably
still of restricted use 
in practice. Here, the reason is that, in general, the second
$\Z[\zeta_{5}]$-direction cannot be chosen in such a way that it
yields dense lines in Penrose model sets $\varLambda_{\text{P}}$, in
other words, one would have to deal with images of poor resolution. A
deeper result is the following, which deals with the class of
convex subsets of a fixed PMS $\varLambda_{\text{P}}$.

\begin{theorem}\label{main1}
There is a set $U\subset \mathbb{S}^1$ of four pairwise non-parallel
$\Z[\zeta_{5}]$-directions such that, for all \textnormal{PMS} $\varLambda_{\textnormal{P}}$, the set
$\mathcal{C}(\varLambda_{\textnormal{P}})$ is determined by the $X$-rays in
the directions of $U$, while, for all \textnormal{PMS}
$\varLambda_{\textnormal{P}}$ and any set $U\subset \mathbb{S}^1$ of
three or less pairwise non-parallel $\Z[\zeta_{5}]$-directions, the set
$\mathcal{C}(\varLambda_{\textnormal{P}})$ is not determined by the $X$-rays in
the directions of $U$. 
\end{theorem}

For example, the set of
$\Z[\zeta_{5}]$-directions parallel to the elements of the following
set $U$ 
has the desired property to determine 
$\mathcal{C}(\varLambda_{\text{P}})$ by the $X$-rays in
its directions, 
\begin{equation}\label{u5}
U:=\left\{(1+\tau)+\zeta_{5},(\tau-1)+\zeta_{5},-\tau+\zeta_{5},2\tau-\zeta_{5}\right\}\,.
\end{equation}

\begin{remark}
By a result of Pleasants~\cite{PABP2}, these directions can yield dense lines in Penrose model sets. It follows that,
 in the practice of quantitative HRTEM,
 the resolution coming from these directions
 is rather high, which makes Theorem~\ref{main1} look promising.
\end{remark}

Above, we restricted the class of finite subsets of a fixed Penrose
model set $\varLambda_{\text{P}}$ under consideration. In order to
obtain positive uniqueness results, a second option is to consider the
interactive technique of successive determination. One has the
following positive results.

\begin{theorem}\label{ccth6a}
If $\varLambda_{\textnormal{P}}$ is a \textnormal{PMS}, the set $\mathcal{F}(\varLambda_{\textnormal{P}})$ is successively
determined by two $X$-rays in $\Z[\zeta_{5}]$-directions, while the set 
$$
\bigcup_{\varLambda_{\textnormal{P}}\mbox{\tiny \,\,\,\textnormal{PMS} }}\mathcal{F}(\varLambda_{\textnormal{P}})
$$
 is successively determined by three $X$-rays in $\Z[\zeta_{5}]$-directions.
\end{theorem}

Unfortunately, this result is again somewhat limited in practice because,
in general, 
one cannot make sure that all the $\Z[\zeta_{5}]$-directions which are used
 match dense lines in Penrose model sets.

\section*{Final Remark}
For further details in this spirit, we refer to~\cite{Gr, GGP,GG,H1} for the
lattice case and~\cite{BG2,H1} for cyclotomic model sets which also
provide a systematic generalization of the setting explained here for
the PMS.

\section*{Acknowledgements}
It is our pleasure to thank U. Grimm, P. Gritzmann, B. Langfeld and
B. Sing for helpful discussions. The authors were supported by the German Research Council
  (Deutsche Forschungsgemeinschaft), within the Collaborative Research
  Centre (Sonderforschungsbereich) 701.

\end{document}